\documentclass[particles,article,accept,pdftex,moreauthors]{Definitions/mdpi}
\firstpage{1}
\pubvolume{1}
\issuenum{1}
\articlenumber{0}
\pubyear{2026}
\copyrightyear{2026}
\externaleditor{Firstname Lastname} 
\datereceived{30 September 2025}
\daterevised{9 January 2026} 
\dateaccepted{9 February 2026}
\datepublished{ }

	\Title{Instruments for Focal Plane X-Ray Polarimetry in the \mbox{Next Decade}}

	\Author{Fabio Muleri $^{1,}$*\orcidA{},
		Stefano Cesare $^{2}$,
		Enrico Costa $^{1}$\orcidB{},
		Walter Cugno {$^{2}$},
		Klaus Desch {$^{3}$}\orcidC{},
		Alessandro Di Marco $^{1}$\orcidD{},
		Sergio Fabiani $^{1}$\orcidE{},
		Riccardo Ferrazzoli $^{1}$\orcidF{},
		Markus Gruber {$^{3}$}\orcidG{},
		Daniel Heuchel $^{4}$\orcidH{},
		Saba Imtiaz $^{1}$\orcidI{},  
		\mbox{Jochen Kaminski {$^{3}$}\orcidJ{},}
		Dawoon Edwin Kim $^{1}$\orcidK{},
		Alessandro Lacerenza $^{1}$,
		Carlo Lefevre $^{1}$\orcidM{},
		Hemanth Manikantan $^{1}$\orcidN{},
		Vladislavs Plesanovs {$^{3}$}\orcidO{},
		John Rankin $^{1}$\orcidP{},
		Ajay Ratheesh $^{5}$\orcidQ{},
		Alda Rubini $^{1}$ and
		Paolo Soffitta $^{1}$\orcidS{}}
	
	\AuthorNames{Fabio Muleri, Stefano Cesare, Enrico Costa, Walter Cugno, Klaus Desch, Alessandro Di Marco, Sergio Fabiani, Riccardo Ferrazzoli, Markus Gruber, Daniel Heuchel, Saba Imtiaz, Jochen Kaminski, Dawoon Edwin Kim, Alessandro Lacerenza, Carlo Lefevre, Hemanth Manikantan, Vladislavs Plesanovs, John Rankin, Ajay Ratheesh, Alda Rubini and Paolo Soffitta}
	\address{
		$^{1}$ \quad {Istituto Nazionale di Astrofisica - Istituto di Astrofisica e Planetologia Spaziali},
		Via del Fosso del Cavaliere, 100, {00133} Rome, Italy;
		{enrico.costa@inaf.it (E.C.); alessandro.dimarco@inaf.it (A.D.M.); sergio.fabiani@inaf.it (S.F.); riccardo.ferrazzoli@inaf.it (R.F.); saba.imtiaz@inaf.it (S.I.); dawoon.kim@inaf.it (D.E.K.); alessandro.lacerenza@inaf.it (A.L.); carlo.lefevre@inaf.it (C.L.); hemanth.manikantan@inaf.it (H.M.); john.rankin@inaf.it (J.R.); alda.rubini@inaf.it~(A.R.); paolo.soffitta@inaf.it (P.S.)}\\
		
		{$^{2}$} \quad Thales Alenia Space Italia S.p.A., Strada Antica di Collegno, 253, {10146} Turin, Italy; stefano.cesare@thalesaleniaspace.com (S.C.); walter.cugno@thalesaleniaspace.com (W.C.)\\
		
		{$^{3}$} \quad Physikalisches Institut, Universit\"{a}t Bonn, Nussallee 12, {53115} Bonn, Germany; klaus.desch@physik.uni-bonn.de (K.D.); gruber@physik.uni-bonn.de (M.G.); kaminski@uni-bonn.de (J.K.); vladislavs.plesanovs@uni-bonn.de (V.P.)\\
		
		$^{4}$ \quad Deutsches Elektronen-Synchrotron DESY, Notkestr. 85, 22607 Hamburg, Germany; {daniel.heuchel@desy.de}\\
		
		$^{5}$ \quad Physical Research Laboratory, Thaltej, Ahmedabad 380009, Gujarat, India; {ajay.ratheesh@inaf.it}
	}
	
	\corres{Correspondence: fabio.muleri@inaf.it; Tel.: +39-06-4993-4565}
	
	\abstract{The successful detection of X-ray polarization from many celestial sources belonging to different classes by the IXPE mission has opened a new window in X-ray astronomy.
		While an impressive number of scientific topics have already been addressed by IXPE, many of them would benefit from a new class of instrumentation that could be launched on a relatively short time scale. In this contribution, we present the development activities of a focal-plane polarimeter whose goal is to extend the energy range of IXPE up to tens of keV, with better sensitivity and lower background. Our design is based on the use of multilayer mirrors and stacked instrumentation, comprising either a low- or medium-energy imaging photoelectric polarimeter and an active Compton polarimeter. Such an approach relies on hardware with flight heritage and---although still under development for the specific application in X-ray polarimetry---it has the potential to answer compelling scientific questions and to soon become competitive from the point of view of feasibility for space applications.}
	
\begin{document}

		\section{Introduction}

		Linear polarization in the emission of celestial sources can originate from different processes, such as scattering in aspherical geometries (e.g., accretion disks) and non-thermal emission processes (e.g., synchrotron radiation)~\citep{Longair2011}.
		These are common occurrences, especially in sources emitting at higher energies, where each process may leave a peculiar signature in the polarization. Therefore, polarization measurements add two precious observables (the degree and angle of polarization) for constraining emission models~\citep{Trippe2014}.
		Moreover, polarization is strongly affected by general relativity and quantum effects during radiative transfer in regions with strong gravitational and/or magnetic fields~\citep{Dovciak2008,Taverna2014}, and is also sensitive to phenomena that could potentially provide evidence for new physics~\citep{Soffitta2024}.
		It is therefore of the utmost importance to perform polarimetry in the energy range where non-thermal processes dominate and celestial sources are brighter.
		Historically, such an energy range has coincided with that of soft X-rays, from $\sim$1 to $\sim$10~keV, and attempts have been carried out since the dawn of X-ray astronomy~\citep{Novick1972}. Nonetheless, detections have been limited to the brightest sources in the X-ray sky for decades~\citep{Weisskopf1976,Feng2020,Long2022}.
		Finally, with the launch in 2021 of the {Imaging X-ray Polarimetry Explorer (IXPE)},
		a NASA--ASI Small Explorer (SMEX-class) mission, the detection of polarization in the energy range 2--8~keV has been enabled for several tens of celestial sources belonging to different classes~\citep{Weisskopf2022,Soffitta2021}.

An impressive number of scientific topics have been addressed thanks to these observations, including constraining the geometry of the corona in X-ray binaries and supermassive black holes~\mbox{\citep{Krawczynski2022,Farinelli2023,Ursini2023}}, investigating particle acceleration in blazars~\citep{Liodakis2022,DiGesu2023}, studying the physical conditions in the atmospheres of magnetars~\citep{Taverna2022}, mapping the magnetic field in pulsar wind nebulae and supernova remnants~\citep{Xie2022,Bucciantini2023,Ferrazzoli2024}, measuring the viewing geometry of X-ray pulsars~\citep{Doroshenko2022}, and probing the emission history of the supermassive black hole at the center of our galaxy~\citep{Marin2023}.

Alongside these successes, IXPE observations are also providing clear directions on how to proceed in building future instrumentation able to fully exploit the diagnostic potential of X-ray polarimetry~\citep{Soffitta2024_2}.
A first aspect to improve is extending the energy range.
This is necessary because the relevant emission processes extend well beyond the IXPE sensitivity interval; while galactic absorption limits, for a significant fraction of celestial sources, the interest at energies below 1--2~keV, a number of highly polarized processes (reflection, inverse Compton, bremsstrahlung, photon emission and propagation in plasma immersed in strong magnetic fields, etc.) dominate over thermal emission above 10~keV.
An instrument working in this energy range could address, better than IXPE or for the first time, topics such as the study of reflection in accreting sources and in the center of the galaxy, non-thermal processes in supernova remnants or magnetars, and cyclotron lines in \mbox{X-ray pulsars.}

Another much-needed improvement is to increase the collecting area.
Polarization from celestial sources is typically a few \% (although polarization as high as 70--80\% has been observed~\citep{Zane2023,Xie2022}), and in this condition the unpolarized fraction of the signal effectively acts as noise in the measurement, adding to the background.
For this reason, polarimetry requires long integration times, which can be impossible to achieve for faint sources unless one has a sufficiently large collecting area.
Moreover, this would also benefit the study of bright yet spectrally variable sources; not surprisingly, IXPE observed cases in which the polarization varied during the observation~\citep{DiGesu2023,Bobrikova2024}, but its relatively small area limited the minimum temporal scales accessible for detecting polarization.
Of course, increasing the collecting area is mainly a matter of larger mirrors, but IXPE also suffered from large dead time and telemetry/memory restrictions that, in the absence of technical improvements, would jeopardize the study of bright sources with a larger telescope.

Other lessons learned from IXPE include (i) the importance of imaging capabilities to resolve extended sources, reduce the background, and identify serendipitous sources in the field of view; (ii) fast and versatile planning of flight operations to pursue the evolution of variable sources occurring on time scales of about one day; and (iii) the great value of simultaneous spectral-imaging observations.

Building on the IXPE heritage, we are developing a suite of instruments able to tackle the new scientific questions left open---or opened---by IXPE, whose status is described below.

\section{Imaging Photoelectric Polarimetry in the 2--30~keV Energy Range with the GridPix}

The polarimeters on board IXPE~\citep{Baldini2021} are based on the GPD design~\citep{Costa2001, Bellazzini2006, Bellazzini2007}.
To measure polarization (only linear polarization, as technologies capable of detecting circular polarization with reasonable efficiency---although available, for example, for magnetic dichroism studies---are not yet sensitive enough to be exploited in X-ray astronomy), GPDs use the photoelectric effect in a gas mixture, usually dimethyl ether (DME).

\begin{figure}[H]

\subfloat[\centering\label{fig:photoelectric}]{\includegraphics[totalheight=5.5cm]{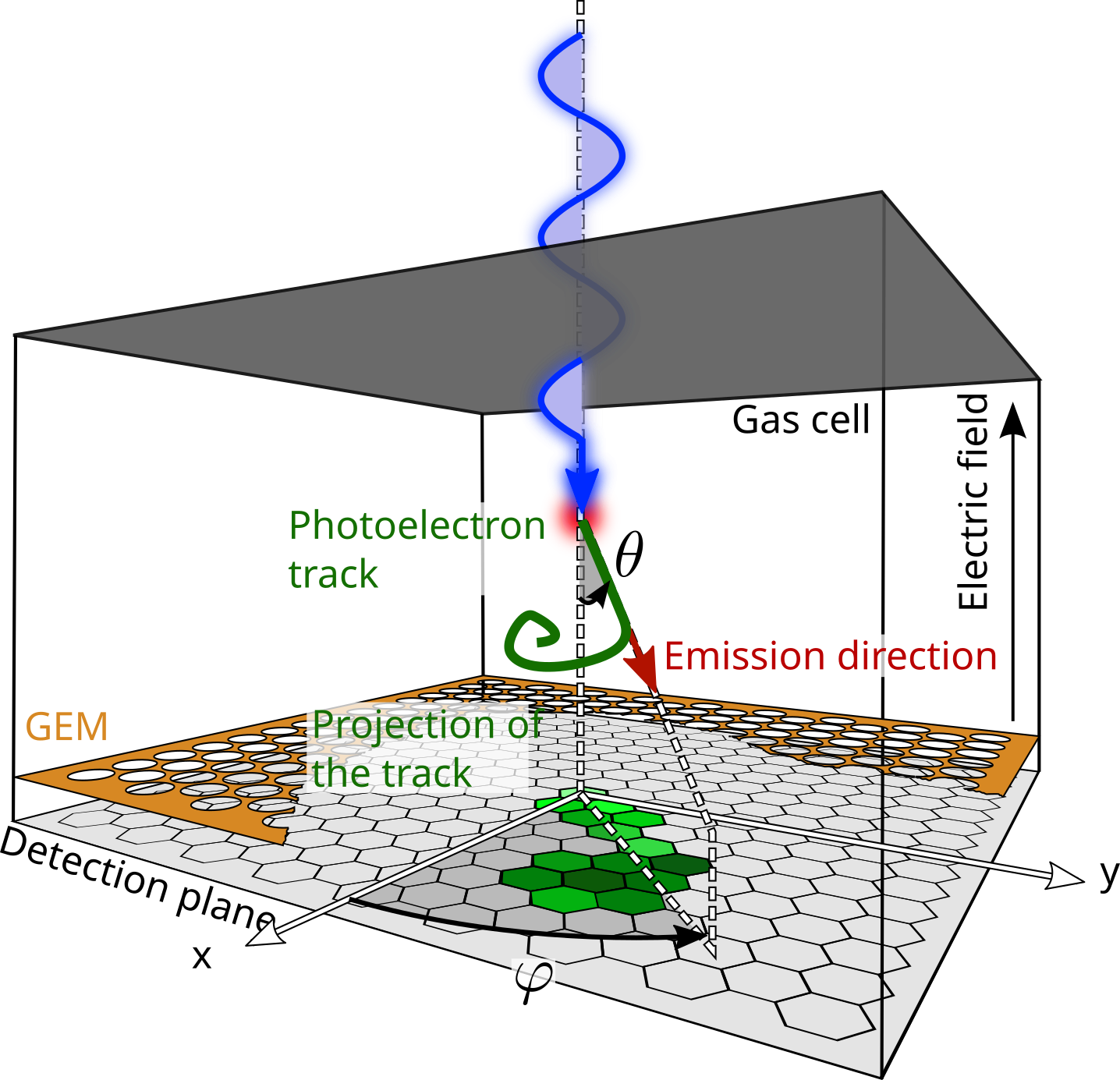}}
\subfloat[\centering\label{fig:ixpe_track_6}]{\includegraphics[totalheight=5.5cm]{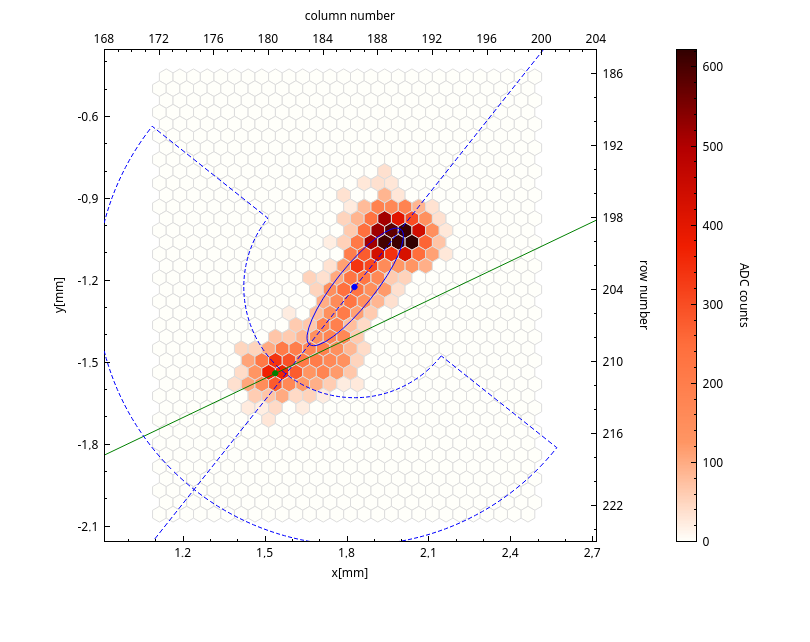}}

\caption{(\textbf{a}) {Operation}
 of a photoelectric polarimeter with the Gas Pixel Detector design. X-rays enter a gas cell passing through a window and are converted into photoelectrons. These propagate in the gas mixture, ionizing atoms along the way. Primary electrons are collected with a drift field, multiplied with a Gas Electron Multiplier (GEM), and eventually imaged by a pixellated ASIC. \mbox{(\textbf{b}) A real} photoelectron track measured by a GPD for 6.4~keV photons. \label{figure1}}
\end{figure}

The operation is sketched in Figure~\ref{figure1}a.
X-rays pass through an optically thin entrance window and are absorbed in the gas, producing a photoelectron.
Its emission direction is correlated with the direction of the electric field of the absorbed photon (with a $\cos^2$ dependency), and therefore this is the quantity to measure for polarimetry.

The photoelectron propagates in the gas, losing energy and producing electron--ion pairs, which are drifted by an electric field. The electrons are multiplied by a Gas Electron Multiplier (GEM) and eventually collected on a pixellated detection plane, which provides a two-dimensional projection of the photoelectron track (see Figure~\ref{figure1}b).

This track is reconstructed to extract the azimuthal emission direction of the photoelectron, whose distribution is expected to be peaked in the direction of polarization~\citep{Muleri2022}.
Moreover, the total collected charge provides the energy of the absorbed photon, the trigger time gives its time of arrival, and the position on the detector provides its incident direction when coupled with an optical system.

Therefore, the GPD measures all the information encoded in a photon, providing true imaging capabilities and a low spurious signal for unpolarized radiation~\citep{Soffitta2021,Rankin2022}.
\vspace{-6pt}

Although the GPD represents the state of the art of photoelectric polarimetry, there is still room for improvement (see \citet{Soffitta2026}).
An effect of charging of the GEM changes the effective gain in the multiplication process during the observation~\citep{Baldini2021}.
The effect depends on the source flux and, to some extent, on its spectrum, and it may cause a change of the energy scale of several \%.
Periodic measurements with calibration sources on board IXPE are necessary to correct for this effect, although some residual miscalibration of the energy scale is often observed ({for example, see}~\citep{Veledina2024}).
Also, the dead time ($\sim$1.3~ms) of IXPE detectors is not negligible in the case of bright sources, although a faster detector has already been developed as an evolution of the one on board IXPE~\citep{Minuti2023}.

To overcome these shortcomings, we are developing a new kind of photoelectric polarimeter based on the GridPix design~\citep{vanderGraaf2007,Kaminski2012}.
This detector was developed by the University of Bonn for the direct search for dark matter through the imaging of recoiling electrons with energies of a few keV. Already used in the CAST experiment~\citep{Krieger2017}, the GridPix will also be the main detector for Baby-IAXO~\citep{Altenmuller2023}.

\begin{figure}[H]
\begin{adjustwidth}{-\extralength}{0cm}
\centering
\subfloat[\centering\label{fig:timepix3}]{\includegraphics[totalheight=5cm]{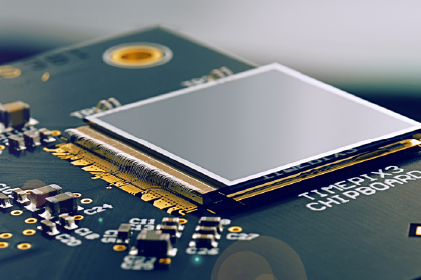}}
\hspace{1mm}
\subfloat[\centering\label{fig:ingrid}]{\includegraphics[totalheight=5cm]{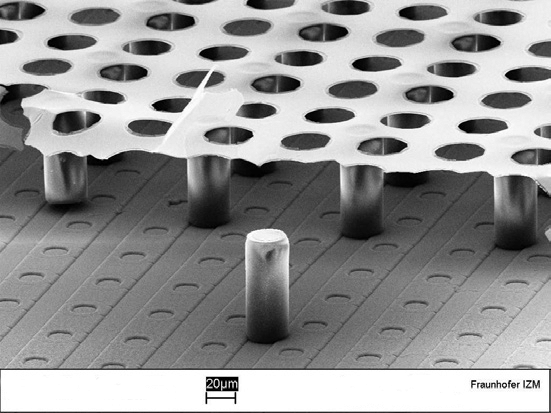}}
\end{adjustwidth}
\caption{{Key}
 technologies of the GridPix detector. Panel (\textbf{a}) shows the Timepix3 ASIC bonded to its read-out board. Panel (\textbf{b}) provides a scanning-electron-microscope view of the InGrid multiplication stage, built directly on the Timepix3. It features a metallic mesh built on top of electrically insulating pillars: electron multiplication occurs in the region delimited by the mesh and the ASIC. \label{fig:gridpix}}
\end{figure}

The GridPix is a gas-flow detector based on two key technologies: the Timepix3 ASIC~\citep{Poikela2014} and the InGrid multiplication stage~\citep{Krieger2013}.
The former, developed by the Medipix collaboration, features 256 $\times$ 256~pixels of size 55 $\times$ 55~{$\upmu$m}
 for a total collecting area of \mbox{14 $\times$ 14~mm$^2$} (see Figure~\ref{fig:gridpix}a).
Pixel noise is 60~e$^-$, and in its first operative mode the chip can measure both the Time over Threshold (ToT) of the signal, which is proportional to the charge collected by the pixel, and its Time of Arrival (ToA) with 1.56~ns resolution.
Such superb timing capabilities imply a maximum rate of 40~MHit/s/cm$^2$, with a very low pixel dead time of 457~ns + ToT.
This complies with the requirements for the observation of celestial sources with a large margin.

The InGrid is a metallic mesh built with photolithography techniques directly on top of the Timepix3 pixels, and separated from them by pillars made of SU-8 that are 50~$\upmu$m high (see Figure~\ref{fig:gridpix}b).
When the InGrid is brought to a voltage $\sim$400--450~V, multiplication occurs in the region between the metallic mesh and the ASIC, which is kept at ground (a protective layer of Si$_3$N$_4$ is deposited on the ASIC surface to protect it from discharges).

The InGrid design reduces the amount of dielectric material that can be subjected to charging effects, and it is therefore expected to solve, or strongly reduce, such effects. Moreover, the small distance between the multiplication and collection regions reduces the diffusion of multiplied electrons, thus increasing the ``sharpness'' of the photoelectron image.
\vspace{-12pt}

\begin{figure}[H]
\begin{adjustwidth}{-\extralength}{0cm}
\centering
\subfloat[\centering\label{fig:tv_setup}]{\includegraphics[totalheight=5.3cm]{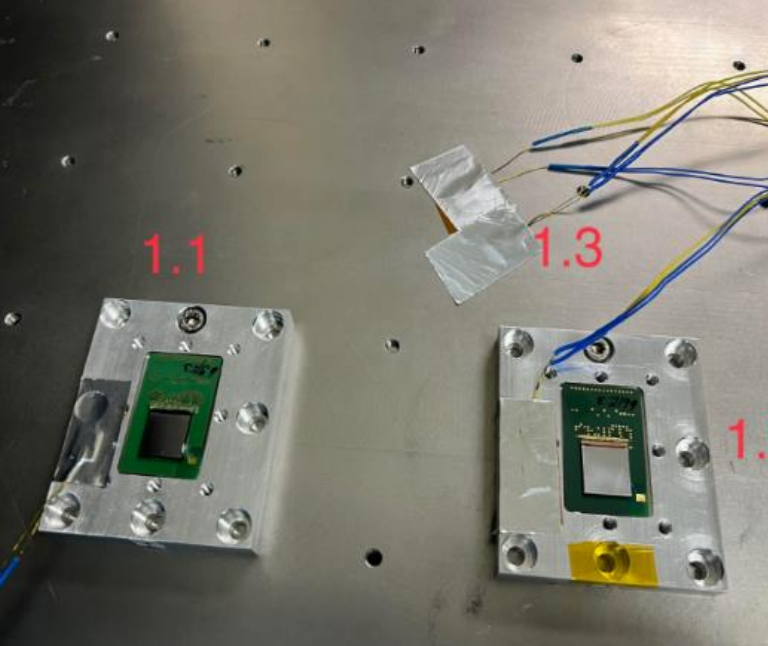}}
\hspace{1mm}
\subfloat[\centering\label{fig:tv_loop}]{\includegraphics[totalheight=5.3cm]{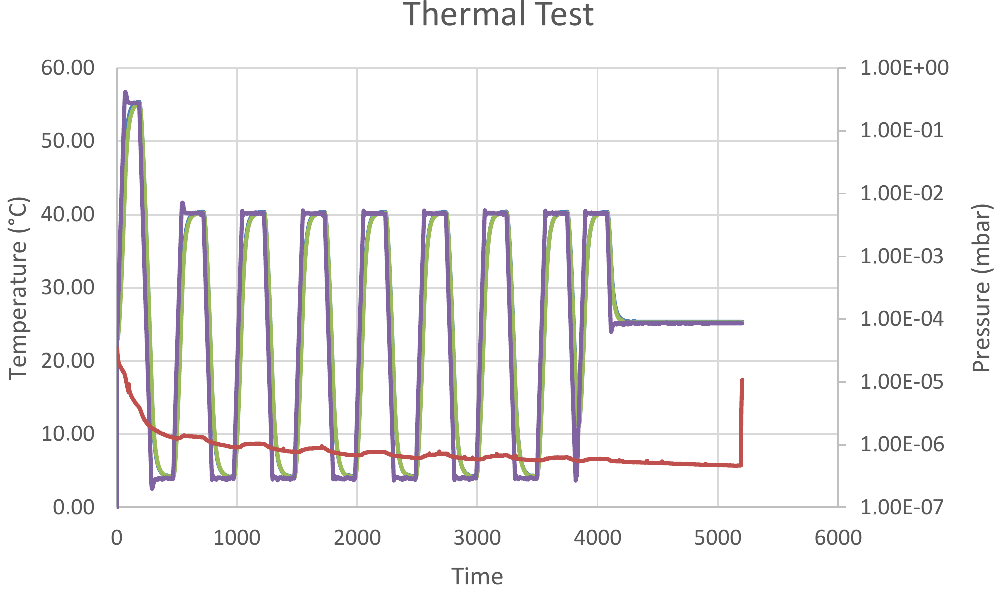}}
\hspace{1mm}
\subfloat[\centering\label{fig:vibration_table}]{\includegraphics[totalheight=5.3cm]{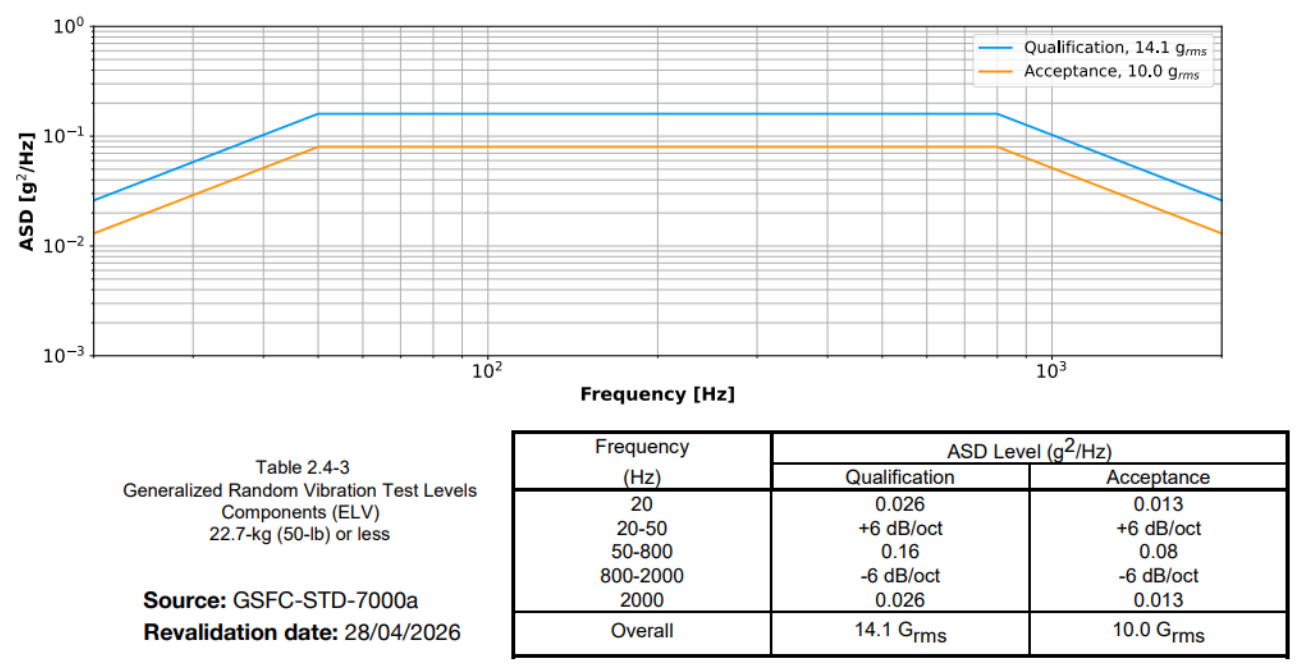}}
\end{adjustwidth}
\caption{{Picture}
 of the set-up for the thermovacuum tests of the InGrid {(\textbf{a})}
 and the temperature as a function of time during the thermal loops {(\textbf{b}).}
 The temperature was cycled from +5 to +40~$^\circ$C, with a peak at 55~$^\circ$C. (\textbf{c}) Vibration profile for the vibration test of the same item. The test was carried out at the qualification level (blue curve), instead of the lower acceptance level (orange line). \label{figure3}}
\end{figure}

A collaboration between the University of Bonn and INAF-IAPS has started to apply the GridPix to X-ray polarimetry of celestial sources~\citep{Ratheesh2024_2}.
The first task was to assess the capability of operating the detector in the harsh space environment.
Timepix3 operation has already been demonstrated in Low-Earth Orbit on board the UKRI SWIMMR-1 and SWIMMR-2 programs.
The InGrid survived thermovacuum cycles between +5~$^\circ$C and 40~$^\circ$C at INAF-IAPS (see Figure~\ref{figure3}b) as well as vibration tests (see Figure~\ref{figure3}c).
Ion-irradiation tests, aimed at verifying the survivability of the detector to heavy ions, are underway at the Isochronous Cyclotron facility of the University of Bonn, and the first results suggest resilience to such events~\citep{Manikantan2025tmp}.
Therefore, the possibility of using the GridPix in the space environment stands on solid premises.

\begin{figure}[H]
\begin{adjustwidth}{-\extralength}{0cm}
\centering

\subfloat[\centering\label{fig:gridpix_setup}]{\includegraphics[totalheight=5cm]{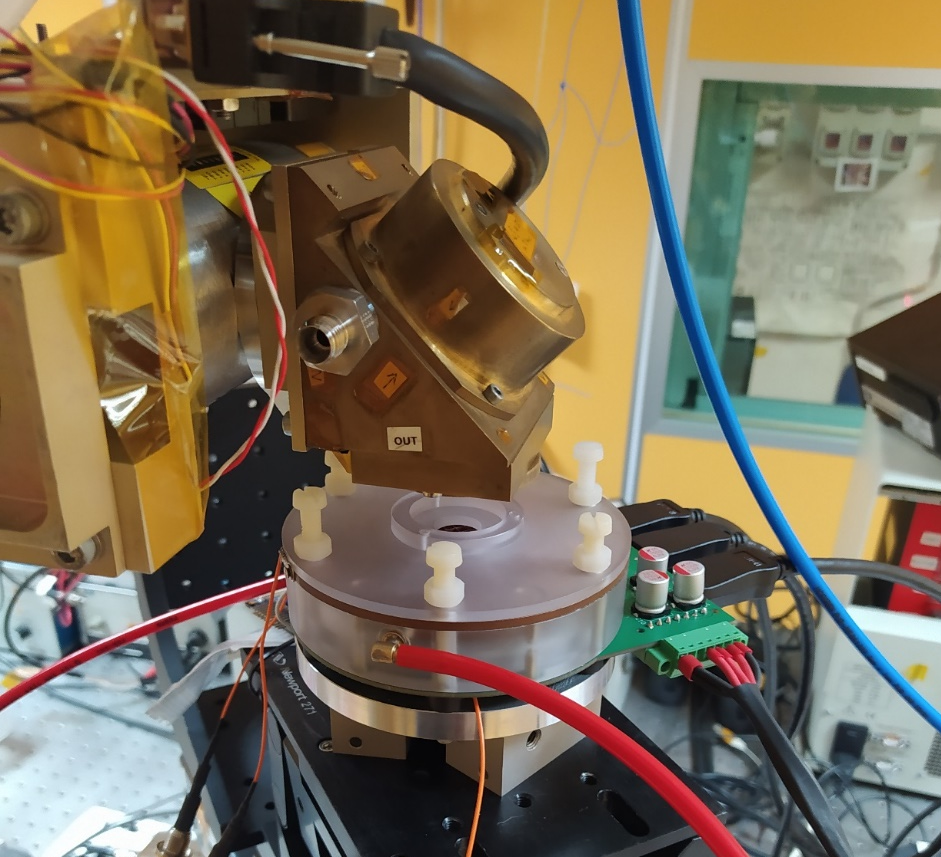}}
\hspace{1mm}
\subfloat[\centering\label{fig:gas_system}]{\includegraphics[totalheight=5cm]{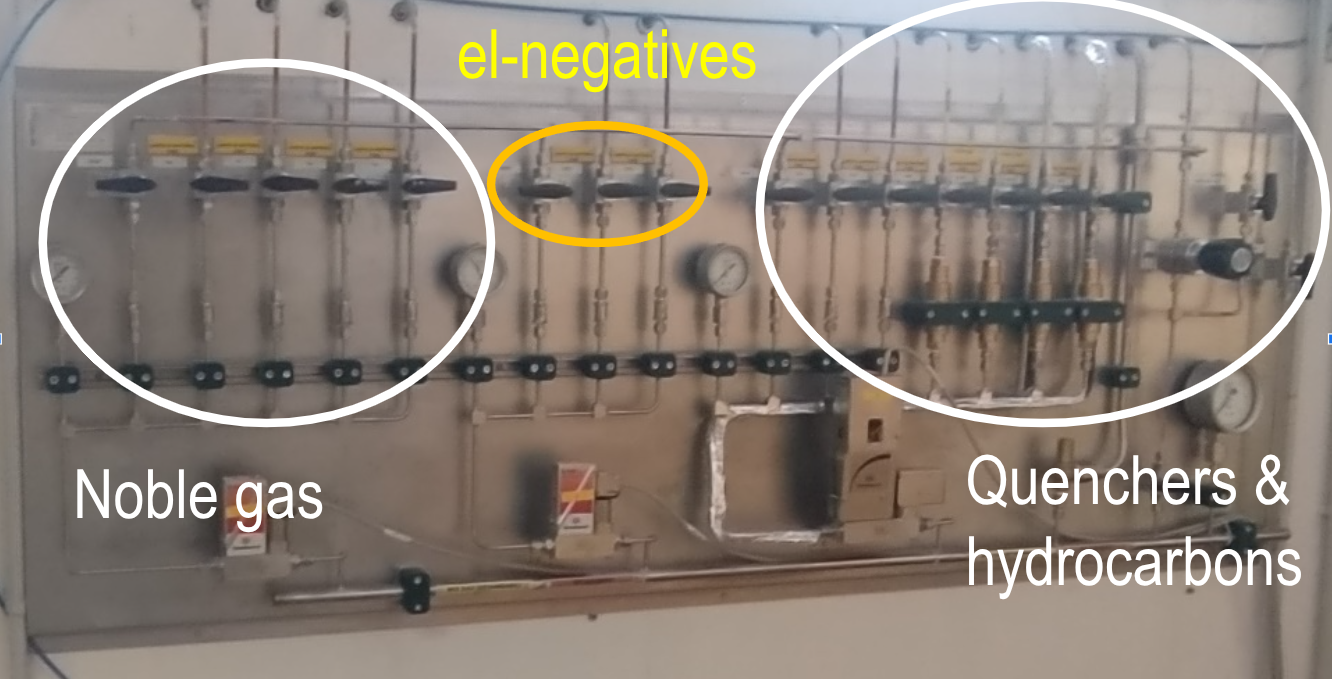}}
\end{adjustwidth}
\caption{(\textbf{a}) The GridPix mounted in one of the INAF-IAPS X-ray test facilities. (\textbf{b}) The gas mixture system at INAF-IAPS used to flow different mixtures inside the GridPix. \label{figure4}}
\end{figure}

\begin{figure}[H]
\begin{adjustwidth}{-\extralength}{0cm}
\centering
\subfloat[\centering\label{fig:gridpix_new_design}]{\includegraphics[totalheight=4cm]{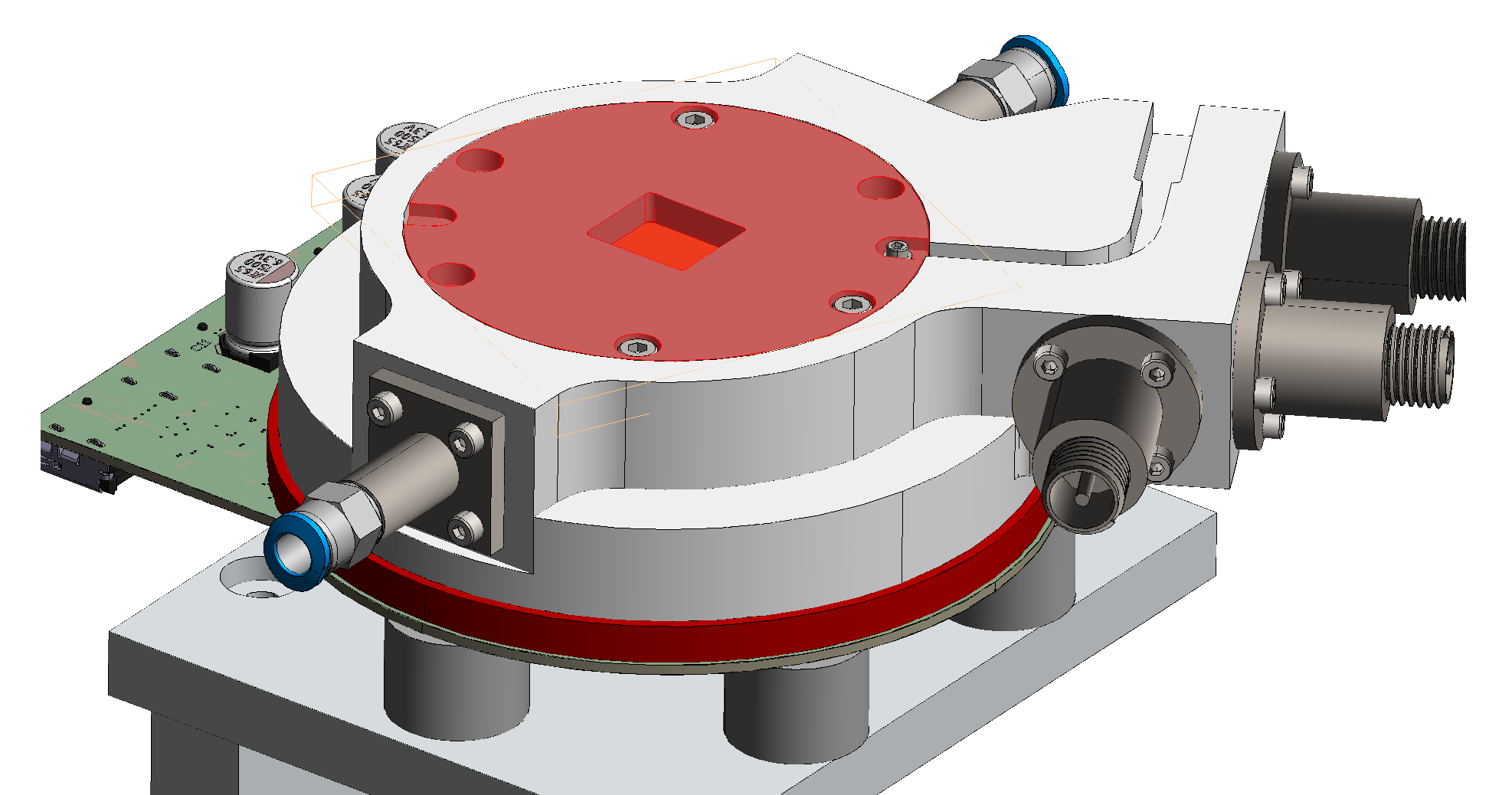}}
\hspace{1mm}
\subfloat[\centering\label{fig:gridpix_new_design_parts}]{\includegraphics[totalheight=4cm]{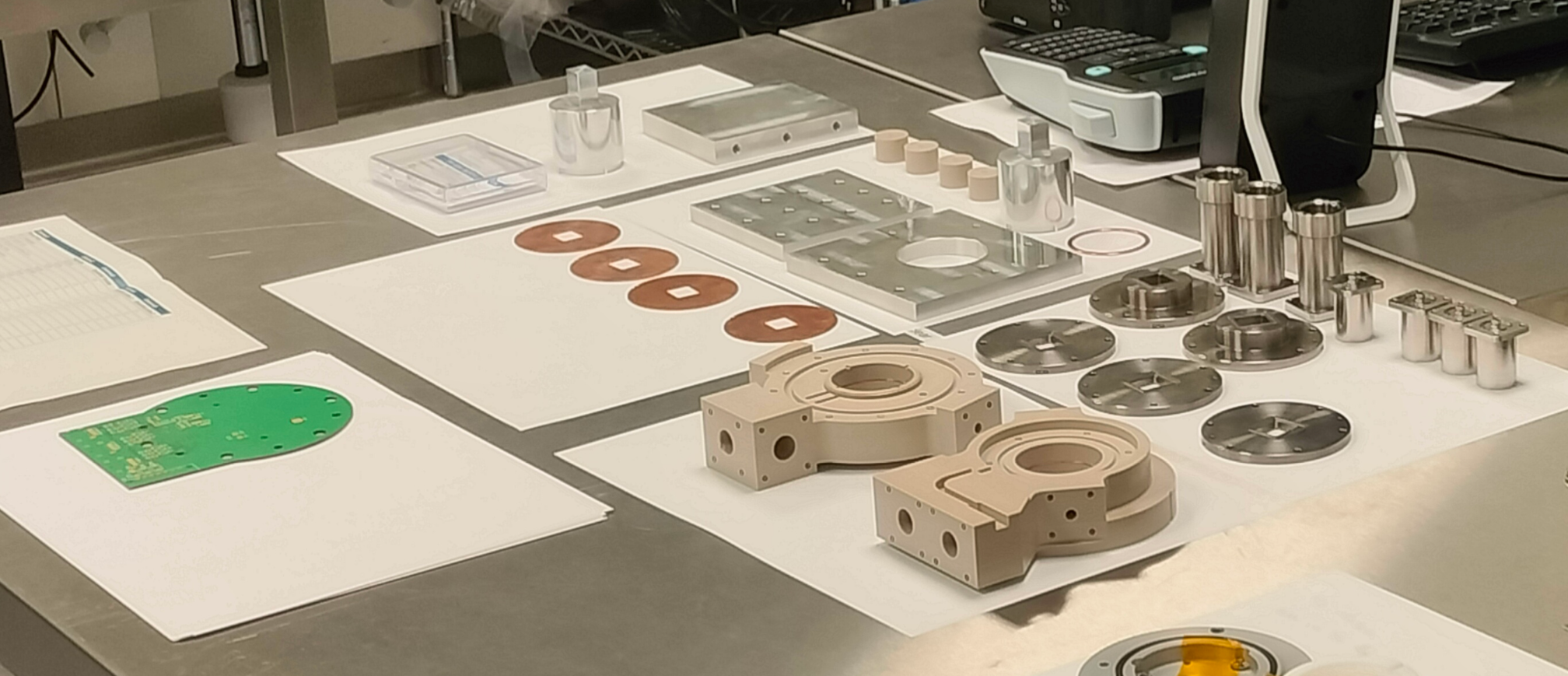}}
\end{adjustwidth}
\caption{(\textbf{a}) The new design optimized for studying the application of the GridPix to X-ray polarimetry. (\textbf{b}) Mechanical parts being assembled in the INAF-IAPS clean room. \label{fig:gridpix_new}}
\end{figure}

In parallel, we started the characterization of the instrument with X-ray sources at the INAF-IAPS test facilities~\citep{Muleri2008,Muleri2022b}.
The detector used is one of those produced for the CAST instrument (see Figure~\ref{figure4}a), which has not yet been optimized for use in X-ray polarimetry (in particular with respect to the uniformity of the drift field).
While a new detector specifically designed for this purpose is being assembled (see Figure~\ref{fig:gridpix_new}), the measurements carried out on the current prototype have already confirmed some of the features that are most interesting for the application to X-ray polarimetry.

\vspace{-12pt}
\unskip

\begin{figure}[H]
\begin{adjustwidth}{-\extralength}{0cm}
\centering
\subfloat[\centering\label{fig:rate_vs_expected}]{\includegraphics[totalheight=5cm]{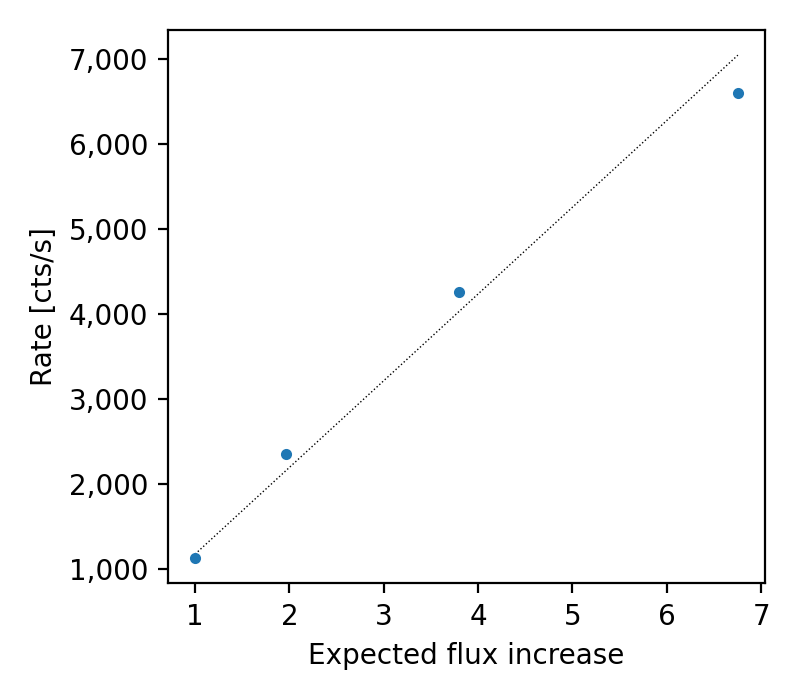}}
\hspace{1mm}
\subfloat[\centering\label{fig:ff8_peak}]{\includegraphics[totalheight=5cm]{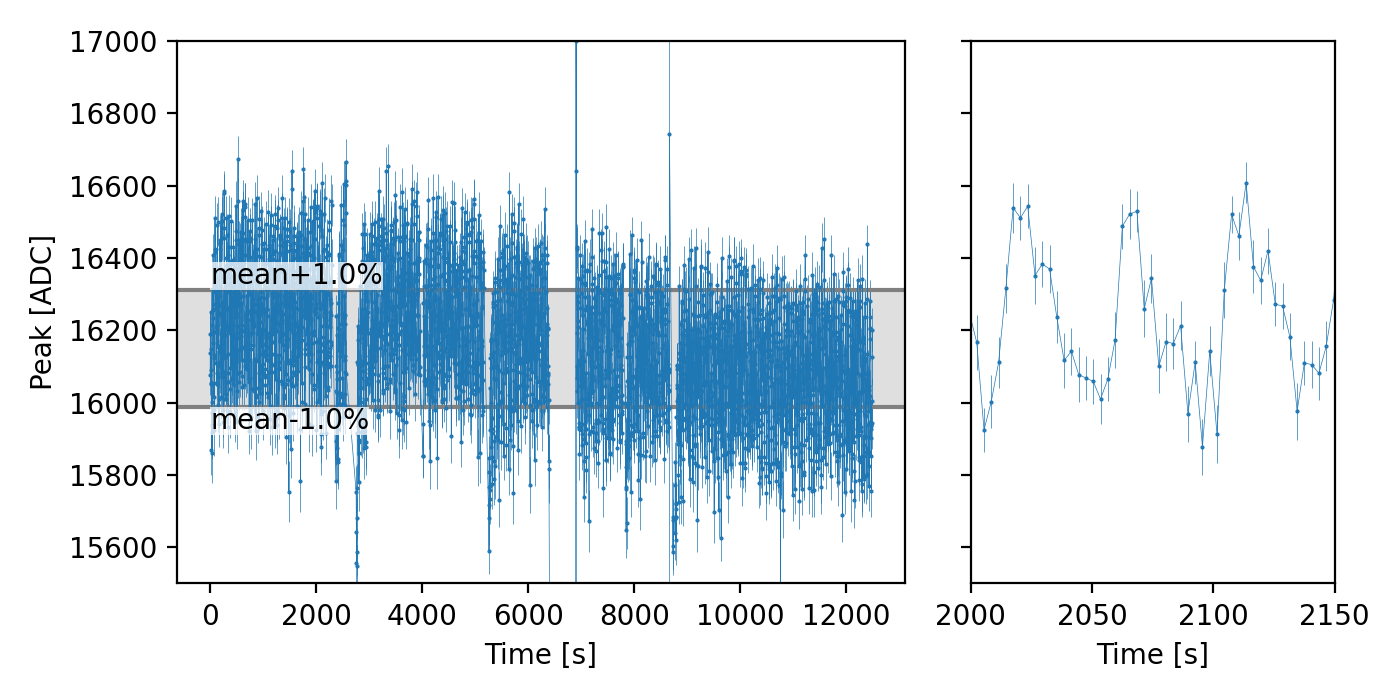}}
\end{adjustwidth}
\caption{(\textbf{a}) {Relation}
 {between the}
 measured rate and the expected flux increase (blue dots) derived by changing in a controlled way the high-voltage and current settings of an X-ray tube with an iron anode. Dashed line represents the expected linear increase. The GridPix was filled with an Argon+CO$_2$ mixture at 1~bar and the gas cell was 1~cm thick. Statistical uncertainties are shown but are not visible. (\textbf{b}) Gain as a function of time measured during an observation with a duration of $\sim$3.5~h. The time bin is 3~s. Gain variation is small, but a periodic change, visible in the zoomed-in view on the right, is present. \label{figure6}}
\end{figure}

We obtained the relation between measured and expected flux by increasing, in a controlled way, the high-voltage and current settings of an X-ray tube with an iron anode, emitting at 6.4~keV.
The relation remains linear at least up to $\sim$7000~counts/s (see Figure~\ref{figure6}a); while a more detailed analysis is underway to better quantify the dead time of the instrument, this result already confirms the absence of dead-time effects up to at least the counting rate that would be measured for a bright source such as the Crab Nebula and for a collecting area two orders of magnitude larger than that of IXPE.

We also started to investigate possible variations of the gain with time with a $3.5$~h long measurement.
The source was an X-ray tube with a copper anode, emitting K$\alpha$ and K$\beta$ lines at 8.0 and 8.7~keV and illuminating the entire sensitive area of the detector.
Data were binned in time and, for each bin, the spectrum was fitted to determine its peak as a function of time, which is shown in Figure~\ref{figure6}b.
The gain appears very stable, with a long-term decrease of $\sim$1\%; however, a modulation of the gain with a similar amplitude is also observed with a periodicity of $\sim$50~s.
Although we are still investigating the origin of such a modulation, the current working hypothesis is that it originates from the power supply used to power the InGrid.

\vspace{-12pt}

A critical feature of the GridPix that was verified during characterization with X-rays was the capability of operating with different gas mixtures.
Taking advantage of the system built at INAF-IAPS that can mix up to three gas components (see Figure~\ref{figure4}b), we verified the operation of the GridPix with helium, neon, or argon mixed with a quencher (CO$_2$ or dimethyl ether), with gas cell thicknesses of 1, 2, or 3~cm.
This demonstrated the capability of the GridPix to operate in an energy range different from the ``classical'' soft X-ray band used by IXPE.
In fact, the energy band of a photoelectric polarimeter is primarily determined by the atomic number of the elements in its gas mixture; by changing the mixture from the DME used in the GPDs on board IXPE to argon-based mixtures, the energy interval shifts from 2--8~keV to 6--$\sim$30~keV~\citep{Muleri2006,Fabiani2012}. Similarly, ``light'' mixtures could operate starting from $\sim$1~keV~\citep{Pacciani2003}.

\begin{figure}[H]
\begin{adjustwidth}{-\extralength}{0cm}
\centering

\subfloat[\centering]{ \includegraphics[totalheight=6cm]{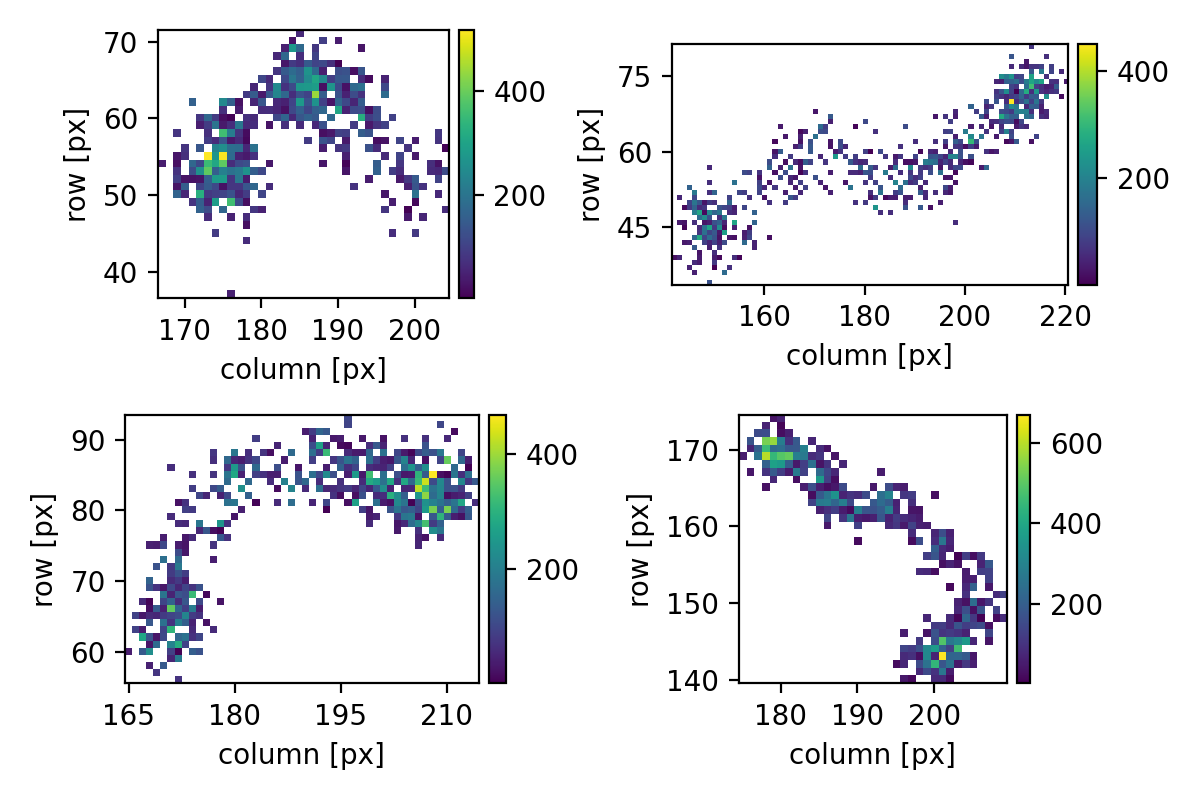}}
\hspace{1mm}
\subfloat[\centering\label{fig:trk130345_recon_3d}]{\includegraphics[totalheight=7.5cm]{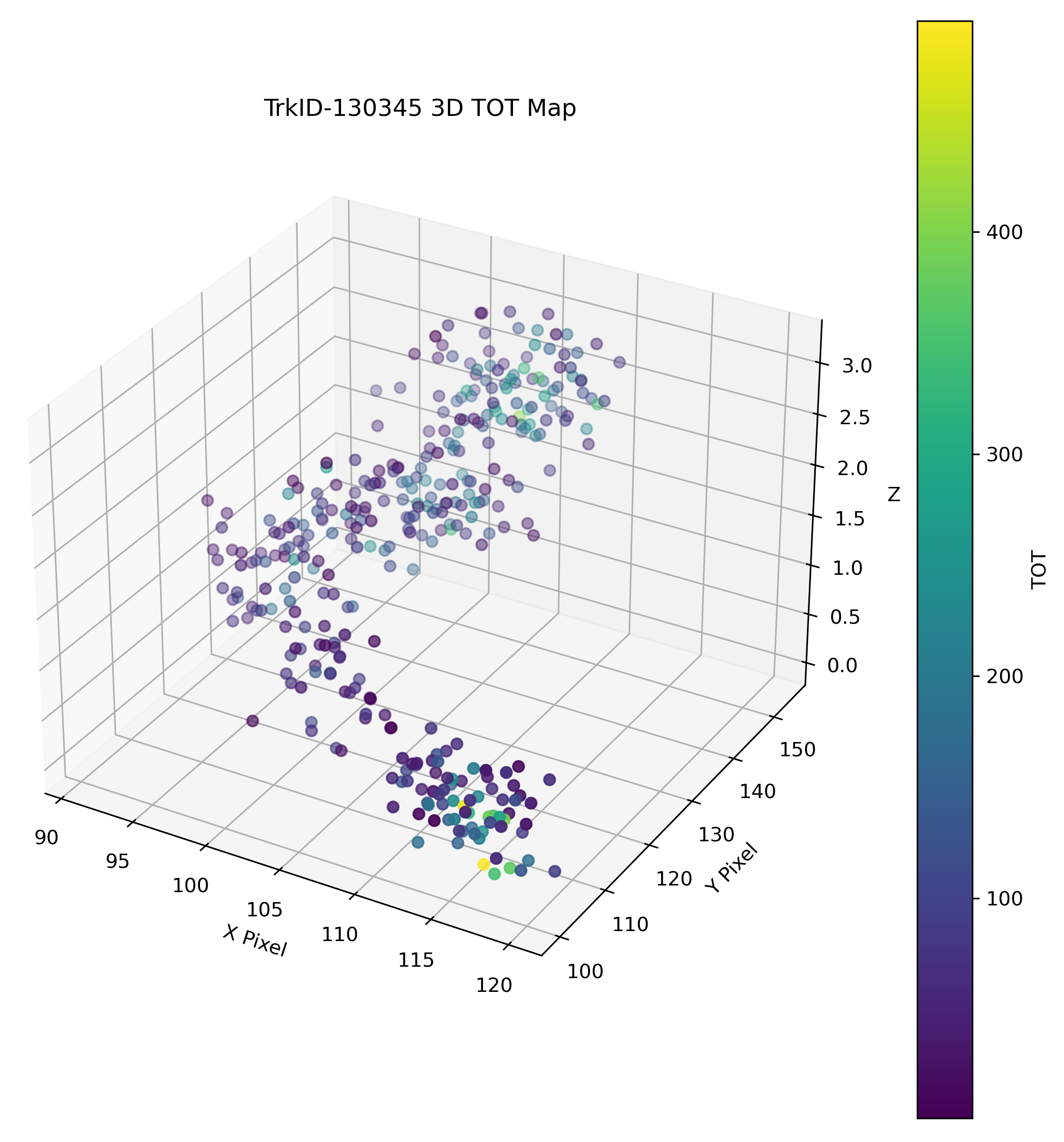}}
\end{adjustwidth}
\caption{Tracks measured with the GridPix at 17.4~keV: ``usual'' two-dimensional projection (\textbf{a}) and an example of three-dimensional reconstruction (\textbf{b}). \label{fig:tracks}}

\end{figure}

Examples of tracks from 17.4~keV photons absorbed in an Ar 80\% + DME 20\% mixture with a 2~cm gas cell are shown in Figure~\ref{fig:tracks}, where we report both the ``classical'' two-dimensional projection, common to other types of photoelectric polarimeters such as the GPD, and a three-dimensional image.
The latter is reconstructed using the unique fast timing capability of the Timepix3; as the pixel signal is resolved with a time resolution of 1.56~ns, this is fast enough to resolve the physical extension of the track along the drift direction.
This information is not strictly required for polarimetry because the distribution of the azimuthal angle does not depend on the polar direction, at least when photons are incident on-axis~\citep{Muleri2014}.
Nonetheless, the reconstruction of the direction benefits from this information in order to distinguish between tracks emitted parallel or orthogonal to the collection plane; the former are easier to reconstruct and, in an advanced analysis, should be weighted more than the latter.
In the end, this is expected to boost the sensitivity and increase the capability to reject tracks generated by the background~\citep{Kim2024_2,DiMarco2024}.
\vspace{-12pt}

\section{Polarimetry Beyond 30~keV}

\begin{figure}[H]
\includegraphics[totalheight=6cm]{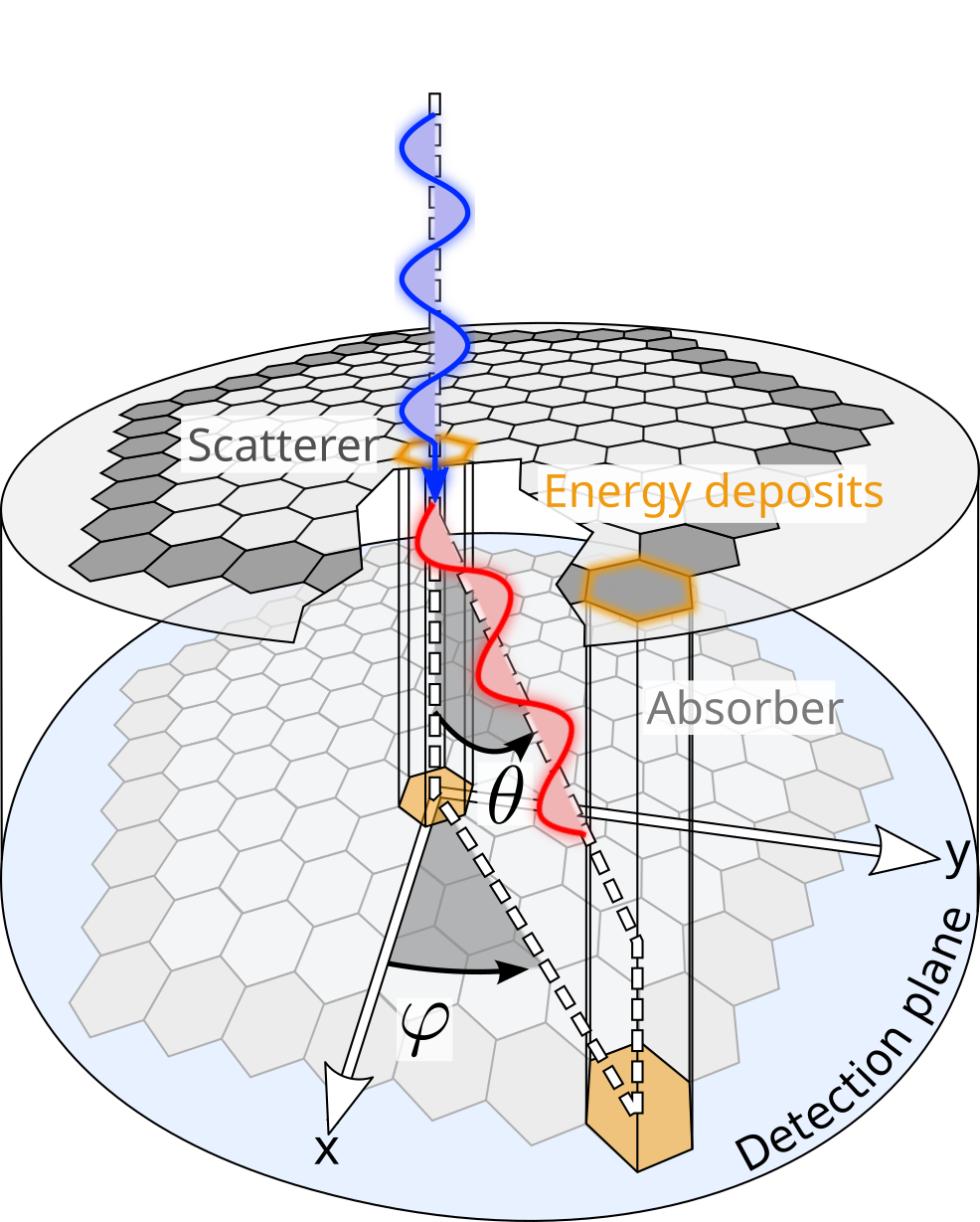}
\caption{{Possible}
implementation of a Compton polarimeter for a focal-plane application. Incident photons (in blue) scatter in a low-atomic-number detector and are then absorbed in a second detector (in red), which provides the azimuthal scattering direction $\varphi$, correlated with the polarization of the incident photon. \label{fig:compton}}
\end{figure}

GridPix detectors filled with appropriate argon-based gas mixtures can operate up to $\sim$30~keV. Beyond this energy, polarization is usually measured using Compton scattering~\citep{DelMonte2023}.
Compton polarimeters have been developed for more than 50~years, for both focal-plane and large-area applications (\citet{Novick1972,Kaaret1994,Produit2024,Fabiani2024}; see also \citet{Fiorina2024, Baracchini2025tmp} for large-area photoelectric polarimeters operating in this energy range), and practical implementations can vary significantly depending on the requirements.
In the focal-plane application discussed here, the most promising design relies on a two-stage detector, one optimized for scattering and the other for absorbing the scattered photon.
A possible design is shown in Figure~\ref{fig:compton}.
Photons are focused onto a central detector with low atomic number (often a plastic scintillator) to maximize the scattering probability, and are then absorbed in a high-Z material (for example, an inorganic crystal). The line connecting the two hit detectors provides the azimuthal scattering direction, which is more probably orthogonal to the polarization direction of the incident photon.
An asymmetry in the distribution of azimuthal angles then reveals the polarization of the incident radiation.
The energy threshold of a Compton polarimeter can be as low as 20~keV in this design~\citep{Fabiani2013}, thus matching well the high-energy limit of photoelectric polarimeters filled with argon mixtures.
Imaging instruments have also been proposed for the polarimetry mission POLARIS for JAXA~\citep{Hayashida2014}, although these devices may have a higher threshold and a more complex design.
\vspace{-12pt}
\unskip

\section{A Mission Concept for Fast Imaging Polarimetry in a Wide Energy~Range}

The specifications and the results already obtained support the use of the GridPix on board a future mission dedicated to extending imaging X-ray polarimetry beyond the capabilities of IXPE.
A suite of GridPix detectors filled with different mixtures could provide sensitive imaging polarimetry in the energy range between $\sim$2 and $\sim$30~keV.
Multilayer mirrors can easily provide sufficient collecting area also beyond this energy; polarization could then be measured using Compton scattering polarimeters.
The scientific goals and a possible early mission configuration achievable with such a mission were outlined in \citet{Soffitta2021_2}.

\begin{figure}[H]

\subfloat[\centering\label{fig:expo_fairing}]{\includegraphics[totalheight=6cm]{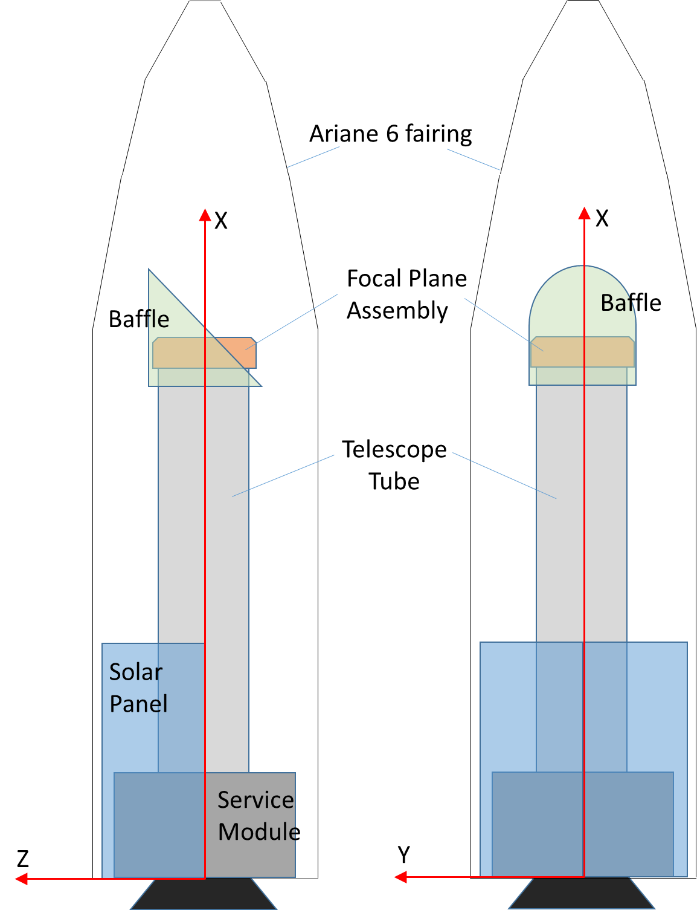}}
\hspace{1mm}
\subfloat[\centering\label{fig:nhxm_area}]{\includegraphics[totalheight=6cm]{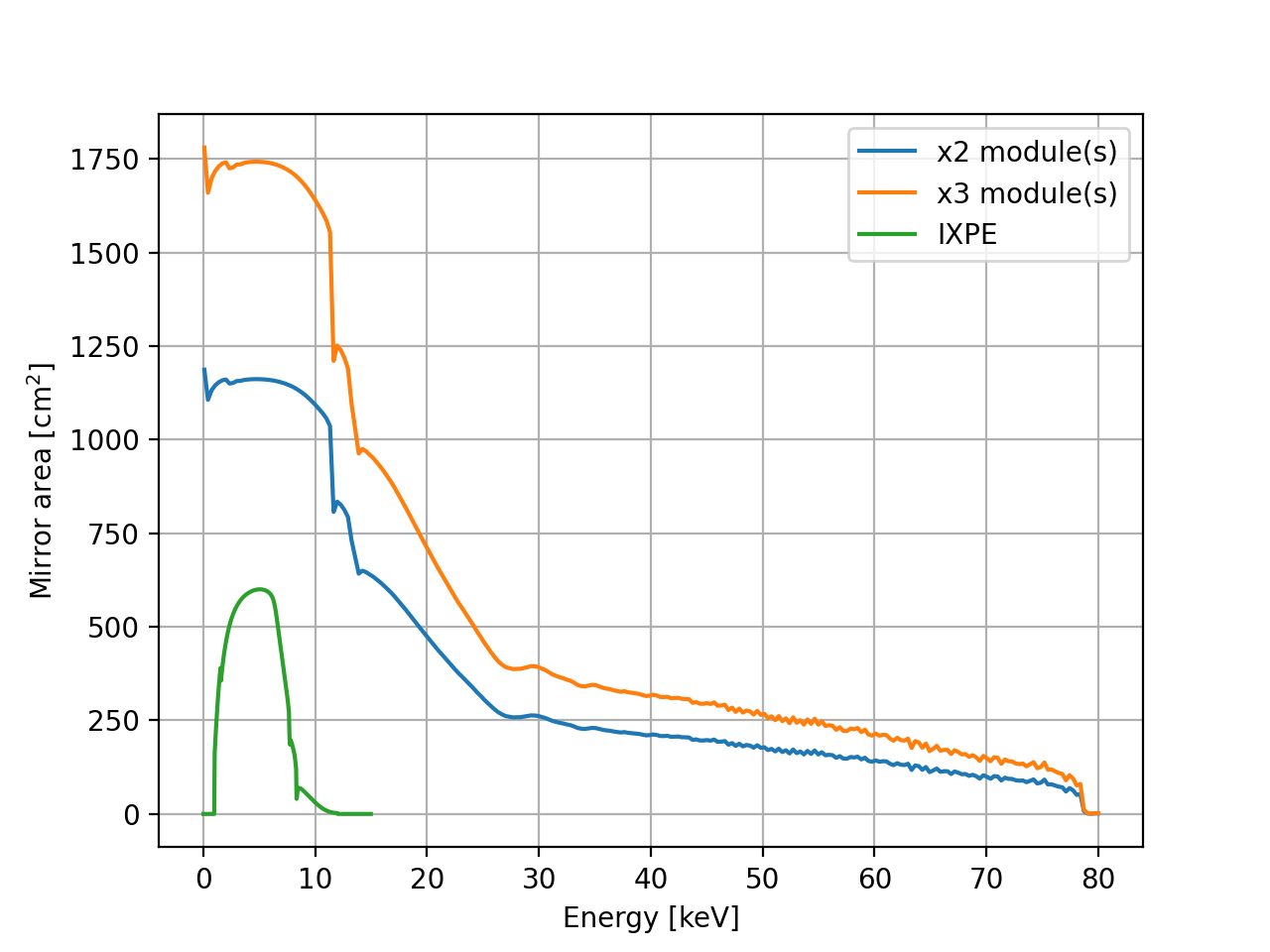}}
\hspace{1mm}
\caption{(\textbf{a}) Accommodation of the assumed mirrors, with a focal length of 10~m, in the fairing of the Ariane~6 launcher. (\textbf{b}) Comparison of the total mirror area with IXPE. Two mirrors are assumed to be dedicated to polarimetry in the soft X-ray band, while three are used in the medium-energy interval to compensate for the typical decrease of flux from celestial sources with energy. \label{figure9}}

\end{figure}

A possible implementation of this mission concept is presented in Table~\ref{tab:payload}.
The payload comprises a small cluster of co-aligned X-ray telescopes with mirrors based on multilayer technology.
To make a concrete example, we assume the design of the mirrors of the New Hard X-ray Mission~\citep{Tagliaferri2012}, which included two polarimeters in the payload covering a wide energy band.
A preliminary evaluation showed that at least six such mirrors, which have a focal length of 10~m, could fit within the fairing of the European launcher Ariane~6 (see Figure~\ref{figure9}a), and we therefore adopt such a configuration as a test case.
The mirrors provide a significant collecting area up to at least 80~keV (see Figure~\ref{figure9}b, \citet{Cotroneo2009}).

\begin{table}[H]

\caption{{A}
 possible configuration for the payload of a mission dedicated to expanding X-ray polarimetry beyond IXPE\label{tab:payload}.}

\begin{tabularx}{\textwidth}{CCCCCCC}
\toprule
  Mirror \# & 1 & 2 & 3 & 4 & 5 & 6 \\
\midrule
  Front & LEP & LEP & MEP & MEP & MEP & \multirow{2}*{SIC}\\
  Behind & HEP & HEP & HEP & HEP & HEP \\
\midrule
  \multicolumn{5}{l}{Low Energy Polarimeters, GridPix Ne/DME filled} & \multicolumn{2}{l}{2--8~keV} \\
  \multicolumn{5}{l}{Medium Energy Polarimeters, GridPix Ar/DME filled} & \multicolumn{2}{l}{6--30~keV} \\
  \multicolumn{5}{l}{High Energy Polarimeters, Compton scattering}  & \multicolumn{2}{l}{20--80~keV} \\
  \multicolumn{5}{l}{Spectrometer-Imager Camera} & \multicolumn{2}{l}{0.3--80~keV} \\
\bottomrule
\end{tabularx}
\end{table}

Five telescopes are dedicated to X-ray polarimetry. A stacked design is implemented to cover the entire energy bandpass of the mirror; an imaging photoelectric polarimeter based on the GridPix, filled with a mixture optimized either for soft X-ray or medium-energy polarimetry, is mounted on top of a Compton polarimeter, which is sensitive to photons at higher energies that can therefore pass through the upper detector. A high transparency, greater than 90\% at 30~keV, can be achieved by thinning the Timepix3 using a process that has already been defined.
\vspace{-12pt}

In total, the payload comprises two imaging photoelectric polarimeters optimized for soft X-rays (2--8~keV), three dedicated to medium energies (6--30~keV), and five Compton polarimeters operating from 20--80~keV.
The sensitivity of such a suite of instruments matches very well across the three different, slightly overlapping energy bands; the Minimum Detectable Polarization~\citep{Weisskopf2010,Muleri2022} for an observation of 100~ks of a source with a flux comparable to that of the Crab Nebula is 0.21\% in the 2--8~keV range with the Low Energy Polarimeters (LEPs), 0.23\% in the 6--30~keV interval with the Medium Energy Polarimeters (MEPs), and 0.25\% from 20--80~keV for the High Energy Polarimeters (HEPs). These values are calculated assuming no spurious modulation and a negligible background; the impact of these two contributions on the sensitivity is currently being evaluated.

To complement the polarimetric capabilities, one mirror hosts a spectral-imaging camera covering the entire energy range of the mirror, and a wide-field X-ray monitor is co-aligned with the other telescopes. The former is intended to provide strictly simultaneous spectroscopy and timing of the same sources observed with the polarimeters, but with $\sim$10~times better energy resolution and higher quantum efficiency.
This would allow a direct relation between the polarization results and the spectral state of the sources.
The X-ray monitor is instead intended to observe a significant fraction of the sky and alert on bursts and spectral or temporal variations of X-ray sources, in order to trigger subsequent observations with the narrow-field instruments.

Preliminary evaluations have indicated that autonomous, \emph{{Swift}
}-like repointing capabilities are feasible, as well as a launch at the First Lagrangian point of the Earth--Sun system, the same location foreseen for the \emph{{NewAthena}} mission. The first feature would allow polarimetry of fast transients, for example magnetar intermediate bursts and gamma-ray bursts, the latter already attempted by IXPE~\citep{Negro2023}.
The second would essentially double the observing time dedicated to scientific observations; including the factor $\sim$2 increase in collecting area with respect to IXPE, our mission design would achieve the same sensitivity as IXPE in one quarter of the observing time, over a much wider energy range.

\section{Conclusions}

IXPE, with its unique capability of measuring polarization in the soft X-ray energy range, is providing outstanding insights into a number of scientific topics, ranging from the geometry of X-ray sources to the physical conditions in their environments.
Next-generation instruments will need to extend the energy range of the measurements in order to fully capture the complexity of the emission processes, provide higher sensitivity, and offer better control of instrumental systematic effects.

To this aim, we are pursuing imaging photoelectric polarimetry with the GridPix. This is a gas imaging detector initially developed for the search for dark matter interactions, which matches well the requirements for X-ray polarimetry from space.
Built on two key technologies, the Timepix3 ASIC and the InGrid multiplication stage, the GridPix offers outstanding speed that enables a three-dimensional reconstruction of the photoelectron track.
An extensive characterization with X-rays has started and supports the use of the GridPix for X-ray polarimetry, and critical components have already been tested to ensure compatibility with the space environment.

Relying on the characteristics of the GridPix, we can outline a possible mission concept to extend the IXPE capabilities to a wide energy band, with improved sensitivity and versatile flight operations.
Such a mission profile has been submitted to ESA in the context of the 8th call for a medium-size mission; at the time of writing, the decision on the possible selection for the presentation of a detailed proposal is still pending.

\vspace{6pt}

\authorcontributions{
{Conceptualization},
 E.C., K.D., S.F., J.K., F.M. and P.S.;
formal analysis, S.C., W.C., A.D.M., R.F., M.G., S.I., D.{E.}
K., H.M., F.M., V.P. and A.R. (Ajay Ratheesh);
funding acquisition, K.D. and P.S.;
investigation, S.F., M.G., D.H., S.I., A.L., H.M., F.M., V.P., {A.R. (Ajay Ratheesh)}, {A.R. (Alda Rubini)} and P.S.;
methodology, J.K., C.L., H.M. and P.S.;
software, A.D.M., M.G., D.{E.}K., H.M., F.M. and J.R.;
supervision, E.C., K.D., J.K. and P.S.;
writing---original draft preparation, F.M.;
writing---review and editing, {all authors.}
All authors have read and agreed to the published version of the manuscript.}

\funding{{We}
 acknowledge the partial contribution of Progetti di Ricerca di Rilevante Interesse
Nazionale del Ministero della Università e della ricerca scientifica (PRIN-MUR) HypeX: High Yield
Polarimetry Experiment in X-rays (Hype-X) prot. 2020MZ884C. Parts of this work have received
funding from the German Federal Ministry of Education and Research under grant no. 05K22PD1. This
work was supported in part by the Italian Ministry of Foreign Affairs and International
Cooperation.}

\dataavailability{{Data reported are available on motivated request.}}

\conflictsofinterest{{The} 
 authors declare no conflicts of interest.}


\newpage
\abbreviations{Abbreviations}{
The following abbreviations are used in this manuscript:
\\

\noindent
\begin{tabular}{@{}ll}
ASI & Agenzia Spaziale Italiana \\
ASIC & Application Specific Integrated Circuit \\
CAST & CERN Axion Solar Telescope \\
DME & Dimethyl Ether \\
GEM & Gas Electron Multiplier \\
GPD & Gas Pixel Detector \\
HEP & High Energy Polarimeter \\
IAPS & Istituto di Astrofisica e Planetologia Spaziali \\
IAXO & International Axion Observatory \\
INAF & Istituto nazionale di astrofisica \\
IXPE & Imaging X-ray Polarimetry Explorer \\
LEP & Low Energy Polarimeter \\
MEP & Medium Energy Polarimeter \\
NASA & National Aeronautics and Space Administration \\
SIC & Spectrometer-Imager Camera \\
SMEX & SMall EXplorer \\
ToA & Time of Arrival \\
ToT & Time over Threshold \\
\end{tabular}
}
\begin{adjustwidth}{-\extralength}{0cm}

\PublishersNote{}
\end{adjustwidth}
\end{document}